# 2-D straw detectors with high rate capability

N.A. Kuchinskiy[1], V.A. Baranov[1], V.N. Duginov[1], F.E. Zyazyulya[2], A.S. Korenchenko[1], A.O. Kolesnikov[1], N.P. Kravchuk[1], S.A. Movchan[1], A.I. Rudenko[1], V.S. Smirnov[1], N.V. Khomutov[1], V.A. Chekhovsky[2], A.S. Lobko[3], O.V. Misevich[3]

[1] Joint Institute for Nuclear Research, Dubna, Russia

[2] National Scientific and Educational Center of Particle and High Energy Physics, Belarusian State University, Minsk, Belarus

[3] Institute for Nuclear Problems, Belarusian State University, Minsk, Belarus

**Annotation**

Precise measurement of straw axial coordinate (along the anode wire) with accuracy compatible with straw radial coordinate determination by drift time measurement and increase of straw detector rate capability by using straw cathode readout instead of anode readout are presented.

PACS: 29.40.Gx.

# Introduction

The detectors which are based on thin wall straws with a diameter from 4 to 10 mm are widely used today as coordinate detectors, for example in the experiments [1-4]. These detectors have following advantages: high spatial resolution, simple structure and low price. In addition the cylindrical geometry allows to have good mechanical characteristics while being of a low mass. Each straw serves as an individual detector and in case of breakdown it doesn't influence the operation of the rest straws.

At the same time a lot of problems still remain unsolved. Thus with the increase of modern accelerator luminosity the particle flux on detectors also increases. Straw detector rate capability is defined by its granularity (straw diameter and anode readout segmentation).

In the individual straw the radial coordinate accuracy is defined by the drift time of primary ionization and cluster step ($\sigma \sim 100$ µm). Axial accuracy along anode wire is defined by readout method and, for example, an accuracy of about few centimeters is achieved for the resistive wire.

We propose a new method of developing a straw detector with high rate capability. The axial coordinate measurement (along the anode wire) is based on reading the induced signals from straw cathode segments ("cathode straw"). Cathode signal readout in the gas-filled detectors is widely used in the experimental physics, however, until the present moment it was not applied for the straw detector. The surface of the straw cathode is segmented into electrically isolated sections with an independent readout. In this case many "cathode straw" detectors are provided and straw detector rate capability increases. Such solution gives us minimum material budget and detector efficiency is not affected by intermediate anode wire supports.

Besides the increase of the detector rate capability, the straw detector segmentation allows us to arrange the measurement of the axial coordinate with accuracy better than 1 mm. This can be achieved by reading of the signal amplitude from the adjacent segments within the 6-7 mm length or by using the "double wedge" method with the segment's length of about 10 cm.

# 1. Increasing straw detector rate capability

A requirement for detector rate capability sharply arises with the increase of luminosity of modern accelerators. It can be achieved either by increasing the rate performance of detector itself or by reducing its size (granularity). Both of these methods have limitations. For example, a straw with a diameter of 4 mm and a length of 100 cm can operate with a maximum rate of about 500 kHz. The practice shows, however, that attempts to reduce the straw tube diameter raise considerable manufacturing-technological problems and the straw average spatial resolution worsens.

The second method consists in electric separation of the anode wire into sections with the individual readouts (Fig. 1a) [5]. This solution was implemented in the ATLAS LHC TRT experiment [6], where the anode wire of the straw is separated into two electrically independent sections by a glass dielectric piece 7 mm long and 0.25 mm in diameter (Fig. 1b). This allowed to realize two independent half-length detectors in one straw and so decrease the rate per a readout channel.

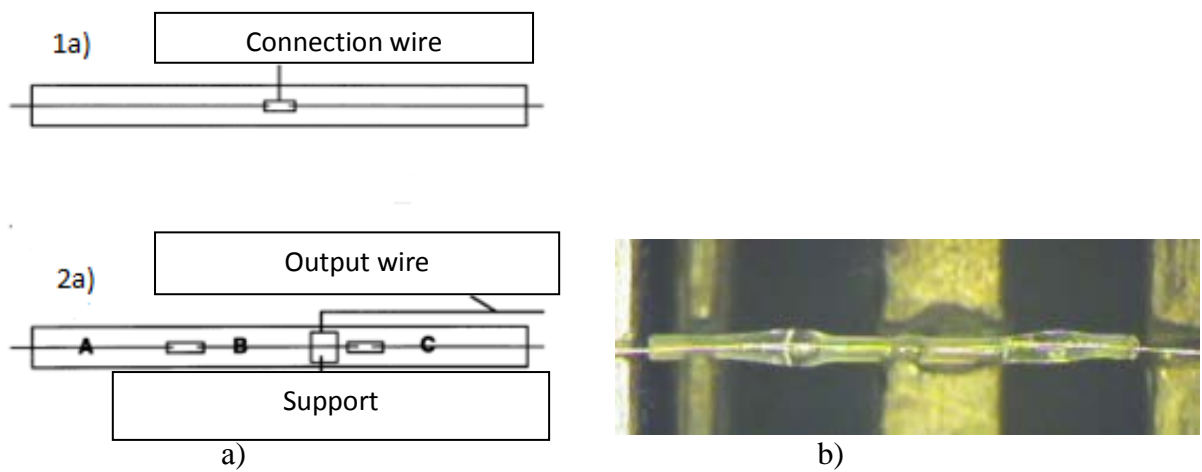

*Fig.1: a) The scheme of separating the straw anode wire into two or three electrically independent sections [5]. b) Glass dielectric piece separating the anode wire (ATLAS, CERN [6]).*

The idea of anode wire segmentation was developed by V. D. Peshekhonov with colleagues [7, 8]. A prototype with multiple segmented anode wires was build (Fig. 2).

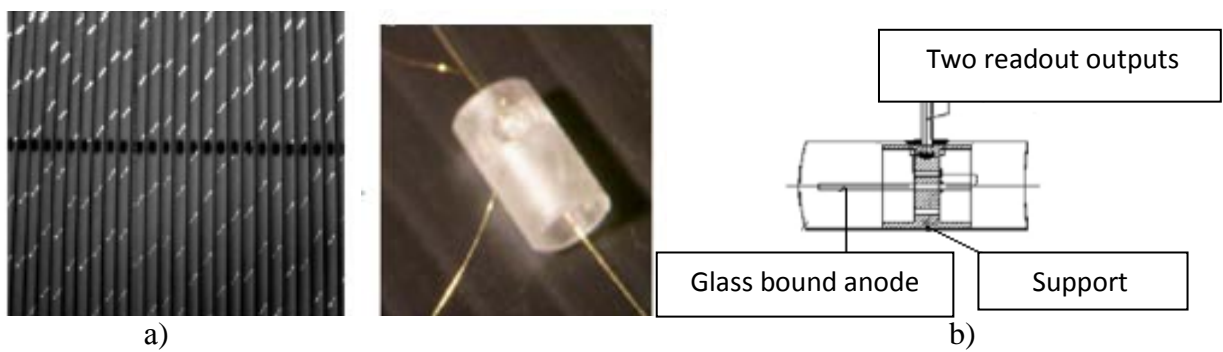

*Fig. 2: a) The fragment of the straw prototype with the segmented anodes [7]. A part with the holes for high voltage feeds and signal output can be seen. b) Anode and spacer.*

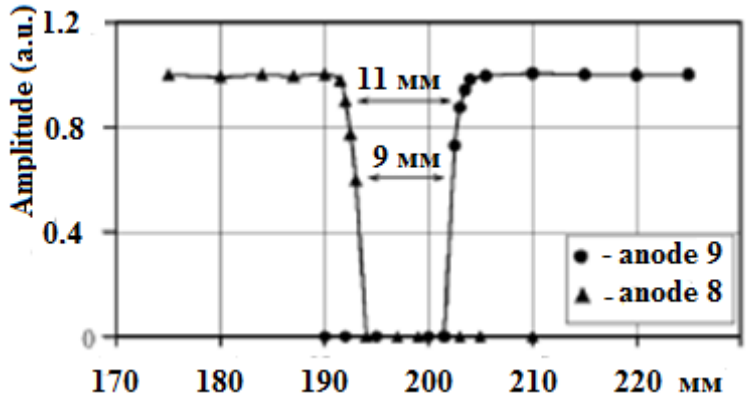

*Fig. 3: Average amplitude of the anode signal between two adjacent anode segments [7].*

The signals from separated sections are brought out through the straw wall with the help of specially constructed spacers (Fig. 2b). The high voltage power supply is connected to anode segments also through these spacers. The length of such anode section, according to the authors' opinion, can reach several centimeters [9]. We should admit that in such method of increasing the straw rate capability one of its main advantages – "transparency" will be lost because of the considerable amount of a material inserted into the working volume of the detector and thus some ineffective zones will appear in the area of the spacers (Fig. 3).

We propose to increase the straw detector rate capability by segmenting the inner surface of the straw cathode into electrically isolated cylindrical sections instead of anode segmentation. The avalanche on the anode wire induces signals on several cathode segments. Signal amplitudes depend on the tube's diameter, the segment's length, and the avalanche position respectively to the line electrically separating the segments. The information from the cathode segments is used both for track radial coordinate measurement according to the drift time value and for the track axial localization according to the segment number or segment charge ratio.

The ability to build the straw detector of such kind depends on the technology of straw manufacturing. About 7000 straws were manufactured by the ultrasonic welding method for NA62 experiment at CERN [10, 11]. Mylar tape 36 µm thick was used to produce the straws 9.9 mm in diameter. Such straws can be used with overpressure up to $dP = 8$ Bar and can operate in the vacuum.

In our measurements we used the straw with a diameter of about 10 mm and a length of 300 mm, which were made of one side metalized 36 µm thick Mylar tape. Double-layered copper/gold metallization is 0.05/0.02 µm thick. The resistance of the conductive coating is 40 Ohm/square. A cross-section of the ultrasonic welding seam is shown in the Fig. 4.

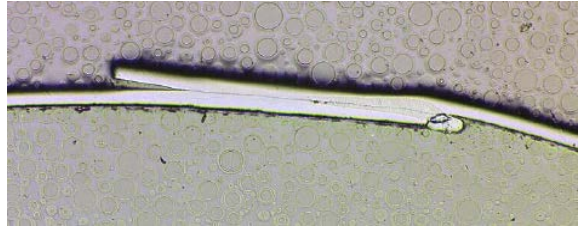

*Fig.4: Cross-section of ultrasonic welding seam.*

Two types of straw were designed and tested: with two and six segments (Fig. 5). The output pads for signals from the cathode segments were provided on the straw ends. The segments and output pads were made on the tape before tape welding.

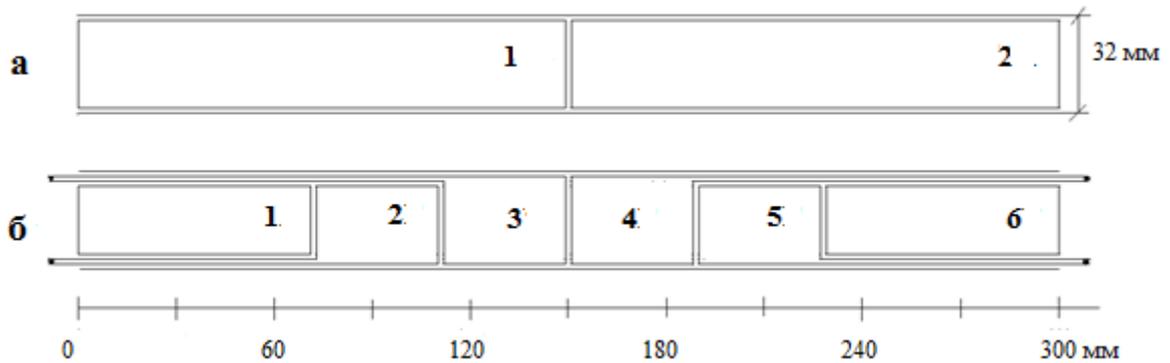

*Fig. 5: Straw with two (a) and six (b) cathode segments.*

The length of the test straw (300 mm) was limited by the size of the test setup. The diameter of the anode wire was 30 µm. Gas mixture of $Ar/CO_2$ (70:30) was used with overpressure dP=10 mBar.

The measurement structure chart is shown in the Fig. 6. The electrons from a $^{90}Sr$ radioactive source irradiate the straw. The straw was moved by a precise mechanical system relatively to the radioactive source and the scintillator counter.

The cathode and anode signals were amplified by CATHODE 1 [12] and Ampl.8.3 [13] circuits and the signal shapes were digitized by a CAEN V1720 (12Bit 250MS/s ADC) [14]. The data were recorded by a computer.

The scintillator counter used for trigger was made of BICRON fiber with a cross-section of 2x2 $mm^2$ and 15 mm long. The light detection was done by a SiPM of CPTA 149-35 type [15] with a sensitive area S=4.41 $mm^2$ and 1764 pixels.

A scan of the $^{90}Sr$ radioactive source along the straw axis was done. The straw segment efficiency vs. straw length dependence is shown in Fig.7. The discriminator threshold value was determined from the cathode segment amplitude spectrum for $^{90}Sr$ radioactive source (Fig. 11). The thresholds were set to 30 mV for each cathode segment.

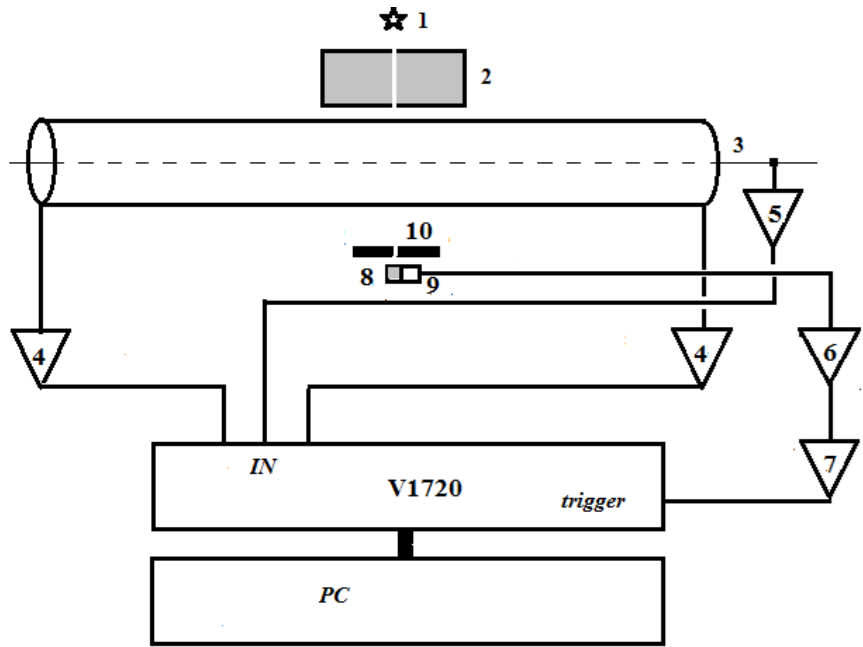

*Fig. 6: The measurement structure chart. 1 - $^{90}$Sr, 2 – flat-field collimator (2 mm), 3 – anode, 4 – cathode amplifiers, 5 –anode amplifier, 6 – SiPM amplifier, 7 – discriminator, 8 – scintillator counter (2x2x5 mm$^3$), 9 – SiPM, 10 – slit collimator (1x5 mm$^2$).*

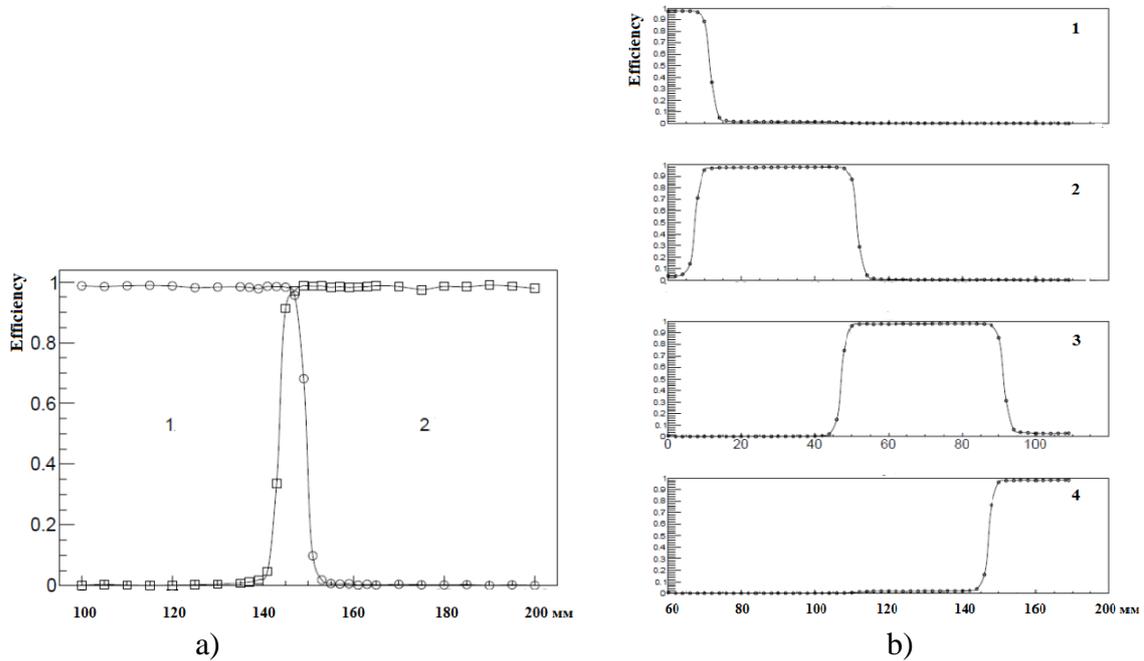

*Fig.7: Straw segment efficiency vs. straw length for a straw: a) with 2 segments; b) with 6 segments (1, 2, 3, 4 – segment numbers).*

The straw cathode segment signal shapes alongside the trigger signal from the scintillator counter are shown in Fig. 8 for the $^{90}$Sr source position near the region of segment separation.

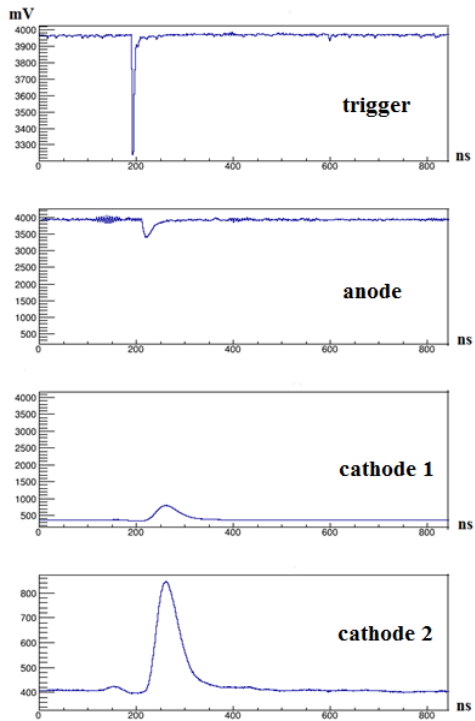

*Fig.8: The straw cathode segment signal shapes with respect to the trigger signal from the scintillator counter. The $^{90}Sr$ source is positioned near the region of segment separation.*

Dependence of the segment efficiency on threshold value and high voltage (HV) value is presented in Fig. 9. The straw HV=3300 V for the threshold scan and the threshold value Th=30 mV for the HV scan.

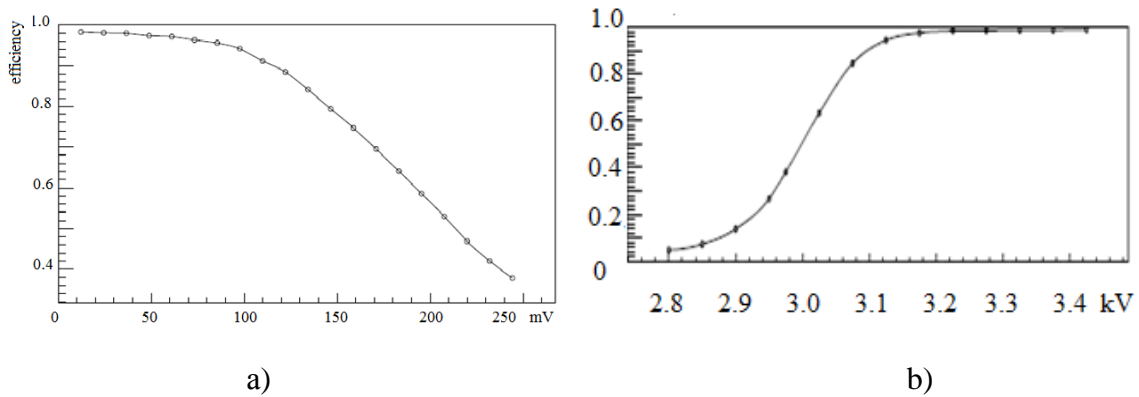

a)                                                                 b)

*Fig.9: Dependence of straw cathode segment efficiency on threshold value (a) at HV=3300 V and dependence of segment efficiency on HV value (b) at threshold Th=30 mV.*

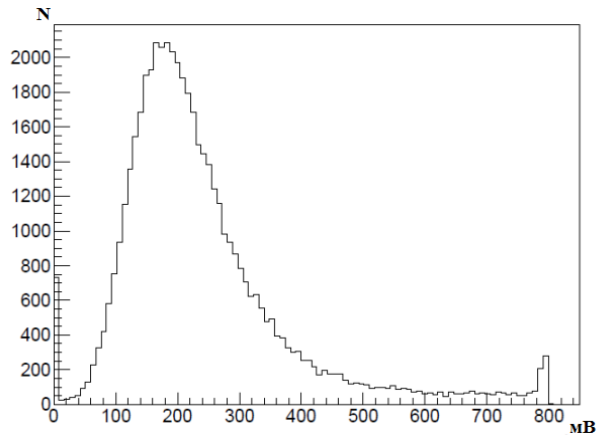

*Fig.10: Typical amplitude spectrum for straw cathode segment.*

The signals from cathode segments can be used for particle track radial coordinate determination by the "drift time measurement" method. Correlation between the "anode drift time" and "cathode segment drift time" is presented in Fig. 11a. Difference between the "segment drift time" and "anode drift time" ($T_c$-$T_a$) is presented in Fig. 11b. The RMS of the distribution is 3.36 ns. Supposing equal contributions of anode and segment signals into common time jitter we can estimate the segment time jitter as dt = 2.4 ns. For a drift speed of about 50 µm/ns we can estimate the coordinate resolution as about 100 µm.

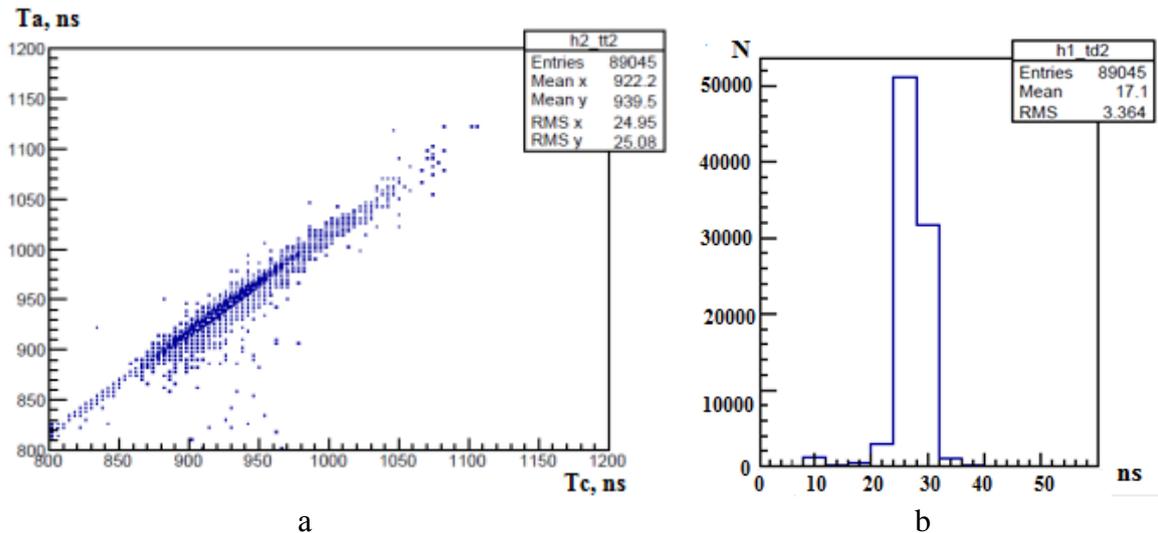

a b

*Fig. 11: Correlation between "anode drift time" and "cathode segment drift time" (a). Difference between "segment drift time" and "anode drift time" (b).*

The ability to read out the signals from straw cathode segments was shown. In case of non-uniform straw irradiation the cathode segments can be made of different lengths in such a way that the rate from each segment will be similar.

As it was mentioned before, in our test straws the signals from the inner cathode segments were delivered to the straw ends by the microstrip lines 1 mm wide. These lines were created on the inner straw metalized surface too. It is obvious that the number of such strips cannot be large. The problem of delivery of the segment signals can be solved if the signals come through the straw wall and go to the straw ends using external straw surface. The holes in the straw wall can be made preliminary before straw welding. The electrical contact between the inner cathode segments and the conductive pads on its external surface can be done, for example, by Kontaktol [16] conductive adhesive glue. After that we have many possibilities to organize the readout from the pads. The precise radial (across the anode wire) and axial (along the anode wire) coordinates of a charged particle track can be measured simultaneously by the same straw.

## 2. Longitudinal coordinate of a straw detector

The second issue is the determination of the straw longitudinal coordinate (along the wire). While the radial coordinate is measured by the method of drift time measurement, determination of the straw longitudinal coordinate (along the anode wire) is a difficult task regarding achievement of precision comparable with the precision of radial coordinate.

At present time the next methods for the longitudinal coordinate measurement are used:

- **method of current division** is a well-known method of the longitudinal coordinate measurement of the avalanche position along the anode wire in the gas-filled detectors. It is based on the comparison of the signal amplitudes read out from the both wire ends [17-20]. This method can also be applied to straw detectors [21, 22]. In order to reach a high precision a wire with high linear resistance is used for the anode, e.g. stainless steel or nichrome. The best spatial resolution has been reached taking into account so called time-charge asymmetry [22]. For example, the spatial resolution is about 1 cm for the straw ends and about 2.5 cm in the center for $^{55}$Fe (2.5 cm and 6.0 cm, respectively, for MIP) at anode wire linear resistance of about 400 Ω/m and the straw length of 1.52 m. The resolution is about (2÷4)% of the straw length.

- **direct measurement of the delay between pulses from the anode wire ends.** This method can also be applied to straw detectors [23-24]. A spatial resolution is about 2 cm for a 2 m straw. A pulse shape digitization unit based on DRS4 (6GHz) chip [25] was used for time measurements [24]. The resolution is about 1% of the straw length. This method is not very promising for short delays due to a strong dependence of the resolution from background noise.

- **charge ratio method for external cathode strips or pads (Fig. 12).** This method is used for straws with a resistive cathode. It allows to achieve a high spatial resolution of about 100 µm along the anode wire. But it is complicated and requires adding of additional readout electrode to the straw detector [26-30].

- **stereo method**. Two crossed layers of straws are used. It allows to reconstruct the longitudinal coordinate of the particle crossing the straws [31,32,33]. The precision of the longitudinal coordinate measurement is $\sigma_z = \sigma_r/\sin(\alpha/2)$, where α is the angle between the straws and $\sigma_r$ is the precision of the radial coordinate measured by drift time. For example, the precision $\sigma_z$ is 2 cm for α = 3° and $\sigma_r$ = 100 µm. It is necessary to have two layers of straws in each inclination angle for high efficiency.

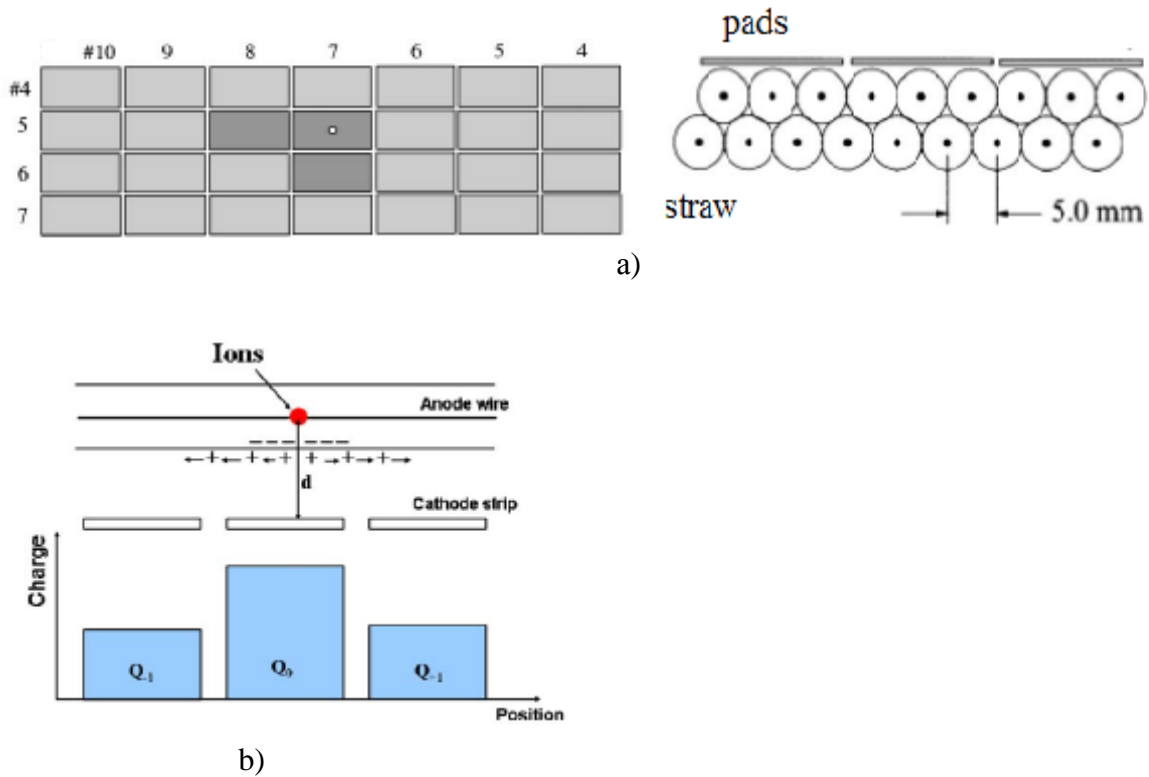

*Fig. 12: a) Pad readout scheme [27]; b) readout with external cathode strips [26A].*

- **"vernier" method**. In 1984 Anderson et al. proposed the so called "vernier" method for the longitudinal coordinate measurement in a rectangular straw [34]. Allison [35] and Green [36] came up with the same idea.

An electrode with double-wedge sequence of cathodes [34] was placed on one of the inner sides of the 11x11 mm² metal square shape straw (Fig.13). An anode wire was stretched in the center. The avalanche coordinate along the anode within one cell with two wedge cathodes was measured by the ratio of the induced charges on them (Q3, Q4). The cell position number along the anode wire was measured by the method of currents division from the two ends of the resistive anode wire (Q1, Q2). The spatial resolution is about 0.8 mm, 1.2 mm and 1.9 mm for a cell length of 10 cm, 20 cm and 30 cm respectively.

This method was used in the chambers for OPAL [37] and MEG [38] experiments.

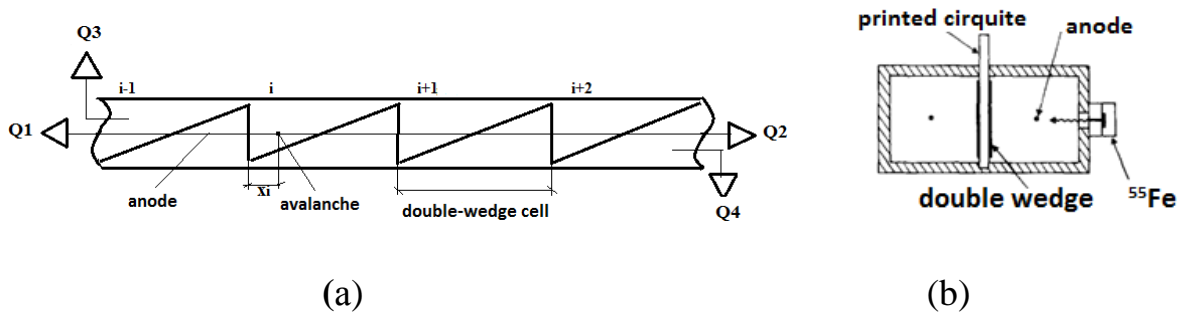

*Fig. 13: Double-wedge printed circuit and measuring diagram are shown (a). Test setup with two square shape straws (b).*

We proposed to use the "vernier" method for the longitudinal coordinate measurement in the straw [39, 40]. It becomes possible due to a new straw manufacturing method based on ultrasonic welding. It is possible to print double-wedge structures on the inner cathode surface of the straw before welding (Fig. 14) [39].

Charge values induced on the cathode pads depend on the avalanche position and it is possible to use their ratio to measure the straw longitudinal coordinate with high precision. The method's accuracy depends on the cell period L (Fig.14.). It should be larger than the spatial resolution given by the current division method based on resistive anode. The spatial resolution is about 0.8 mm for the double-wedge structure with the half period L/2=40 mm and anode wire diameter D=20 µm [39].

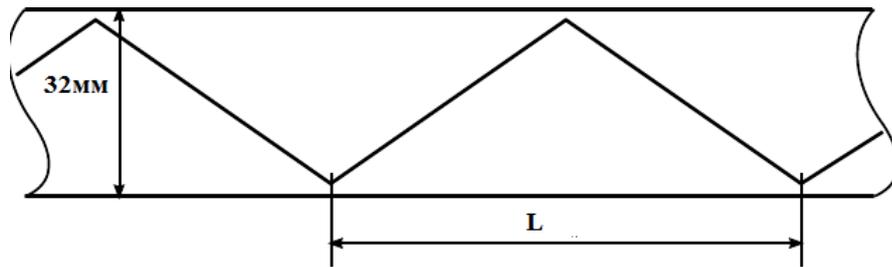

*Fig.14: Tape with a "drawing" double-wedge structure for straw manufacture.*

This method cannot be used at high rates and has limitation in the corners of the zigzag line on the cathode surface due to ambiguity of the solution. It requires additional layers of straws or more than two cathode strips (the double-wedge structure consists of 2 strips).

At the same time the double-wedge method has additional advantage coming from cylindrical symmetry of the induced signal on the cathode surface. Equidistance of the cathode double-wedge structure from the anode wire allows to achieve higher precision compared with a similar detector having planar geometry. The charge ratio of the cathode strips does not depend on the anode signal value.

Straws with a double-wedge structure and a length of 80 mm, 100 mm and 300 mm, anode wire diameter of 30 µm and 20 µm have been also produced and tested. All measurements have been carried out on the test bench described above (see Fig. 5).

The measurement results for the straw with the double-wedge structure period L=80 mm and the anode wire diameter D=20 µm are shown on Fig. 15-17. Ratio of the one strip induced charge Q1 to the total induced charge Q1+Q2 vs. radioactive source position along the straw is shown in Fig. 15. Amplitude spectra of cathode signals from one double-wedge strip for three high voltage values HV=2650 V (1), HV=2725 V (2) and HV=2800 V (3) are shown in Fig. 16. We can see a big difference in the shape of the spectra. Nevertheless the ratio of Q1/Q1+Q2 as function of the HV is constant (Fig. 17). So, the longitudinal coordinate measurement is independent from the gas gain in a straw.

The main disadvantage of the double-wedge structure technique is its low rate capability.

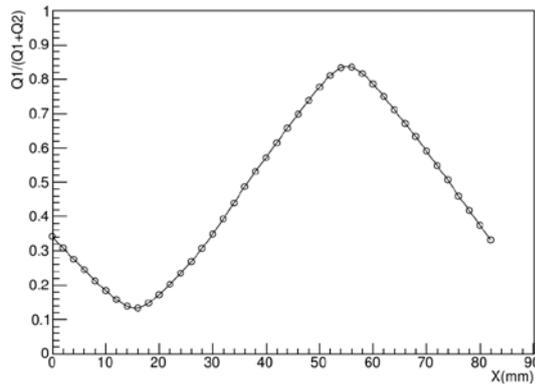

*Fig. 15: Ratio of the one strip induced charge Q1 to the total induced charge Q1+Q2 as function of radioactive source position along the straw.*

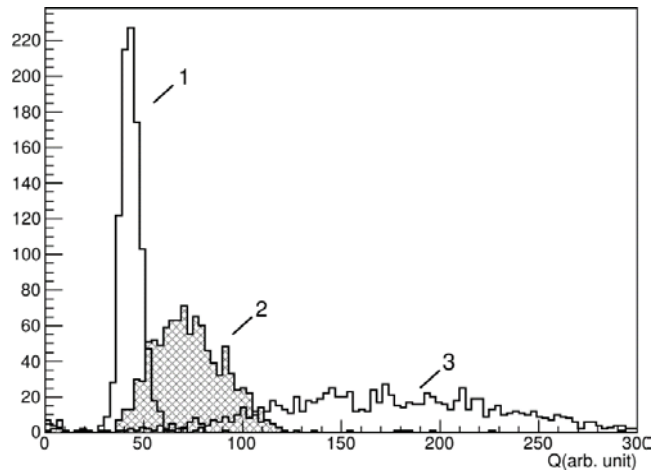

*Fig. 16: Amplitude spectra of cathode signals for three high voltage values HV=2650 V (1), HV=2725 V (2) and HV=2800 V (3.)*

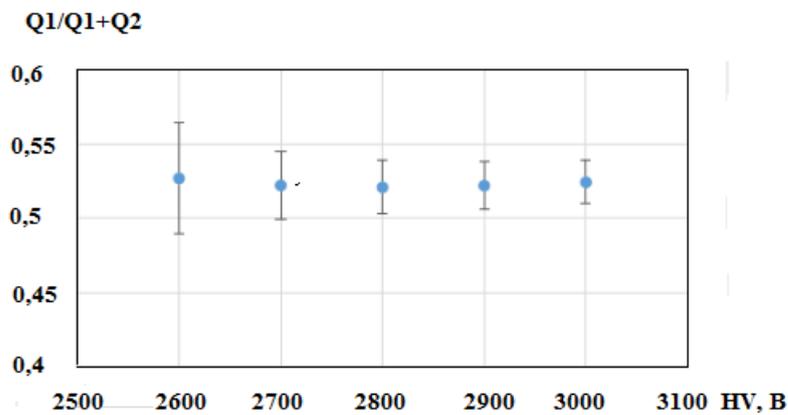

*Fig. 17: Charge ratio Q1/ Q1+Q2 as function of HV value.*

For increasing straw detector rate performance it was proposed to divide a straw into electrically independent segments and apply double wedge method to each segment. Such solution provides independent precise measurement of the longitudinal coordinate within each segment of the straw with high rate performance. The longitudinal coordinate can be measured by the ratio method of induced charges on the neighboring segments.

Straw with cathode segments L=100 mm electrically separated from each other and double-wedge structure inside each segment was manufactured (Fig. 18). Ratio of the charge induced in the segment **2-1** to the total charge in segments **2-1** and **2-2** vs. radioactive source position along the straw is shown in Fig. 19a. Estimated spatial resolution is better than 1 mm (Fig. 19b).

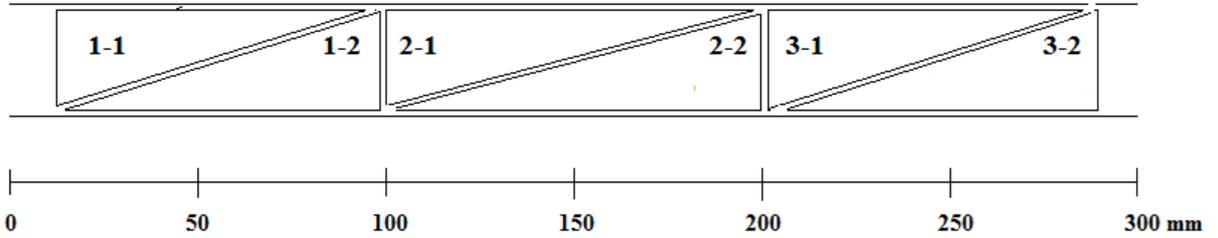

*Fig. 18 Straw with electrically independent cathode segments and double wedge structure in each segment.*

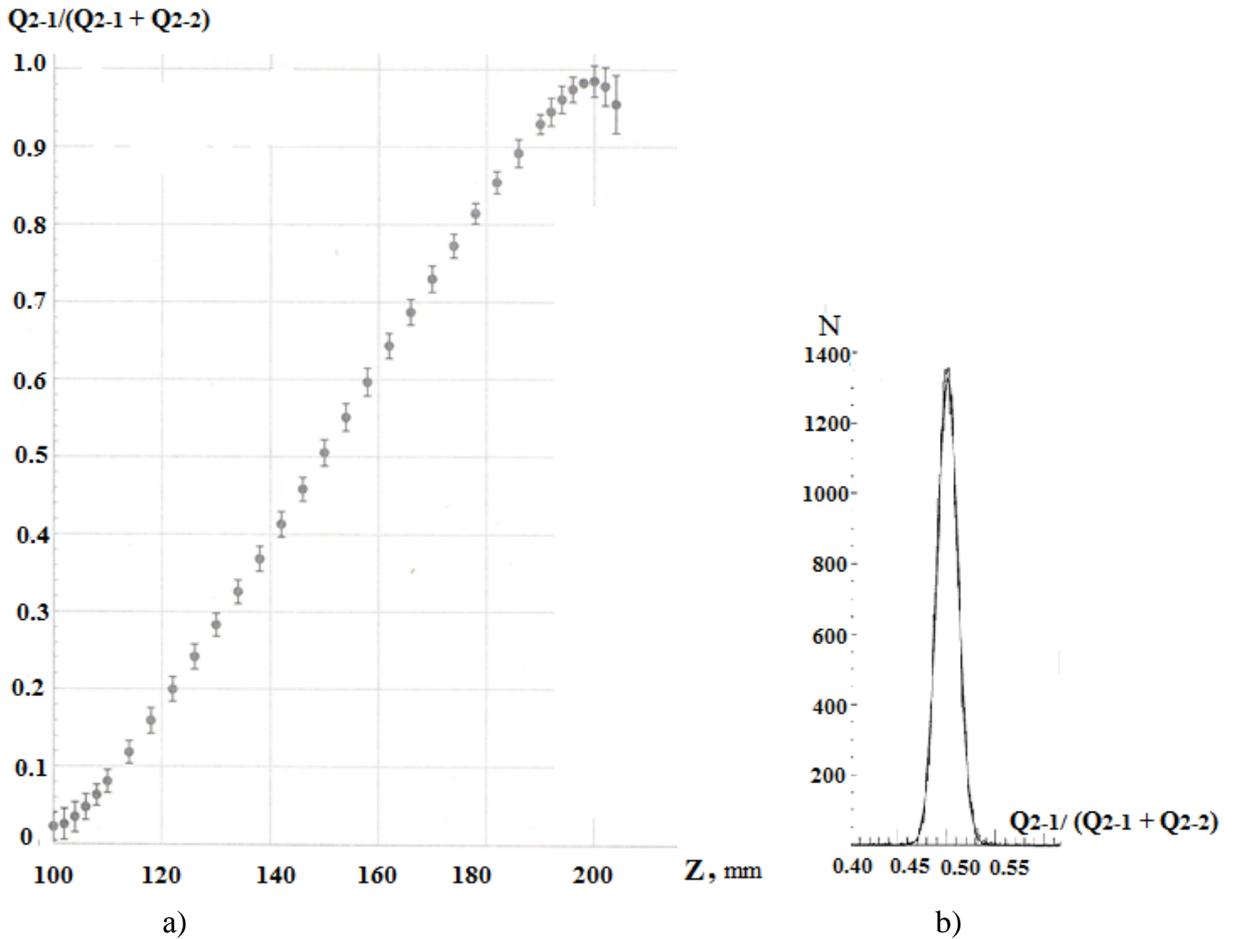

*Fig. 19: a) Ratio of the charge induced in the segment **2-1** to the total charge in segments **2-1** and **2-2** vs. radioactive source position along the straw; b) ratio of the charge induced in segment **2-1** to the total charge in segments **2-1** and **2-2** for collimated radioactive source at a fixed position.*

For a precise measurement of the longitudinal coordinate with the spatial resolution of about 0.1 mm it is necessary to reduce the size of a cathode segment to a value when the induced charge from the anode avalanche reaches three segments simultaneously. It is about 6 mm for a straw with a diameter of 10 mm. The number of segments significantly increases. In this case the signals from the segments can be brought to the straw external surface to simplify its readout by strip or pad readout scheme [40, 41].

Thus, we proposed and demonstrated the possibility to produce the straws with high rate performance and the spatial resolution of about 1% of the segment length (the resolution is better than 1 mm at L=100 mm) for the straw longitudinal coordinate measurement.

This work is supported by Russian Foundation for Basic Research (grant 13-02-00745a) and by Belarusian Republican Foundation for Fundamental Research (grants F13D-006 and F14-003).